\documentclass[prb,preprint]{revtex4}

\usepackage{amsmath}
\usepackage{graphicx}
\usepackage{epstopdf}
\usepackage{color}

\begin{document}
\title{Probability, geometry, and dynamics in the toss of a thick coin}

\author{Ee Hou Yong}
\affiliation{Department of Physics, Harvard University, Cambridge, Massachusetts 02138}
\author{L. Mahadevan}
\affiliation{School of Engineering and Applied Sciences and Department of Physics, Harvard University, Cambridge, Massachusetts 02138}


\begin{abstract}
When a thick cylindrical coin is tossed in the air and lands without bouncing on an inelastic substrate, it ends up on its face or its side. We account for the rigid body dynamics of spin and precession and calculate the probability distribution of heads, tails, and sides for a thick coin as a function of its dimensions and the distribution of its initial conditions. Our theory yields a simple expression for the aspect ratio of homogeneous coins with a prescribed frequency of heads/tails compared to sides, which we validate by tossing experiments using coins of different aspect ratios.
\end{abstract}

\maketitle

\section{Introduction}

Physical problems that involve probabilistic outcomes range from the statistical mechanics of large ensembles of particles to the seemingly simple games of chance such as the toss of a coin and the spin of a roulette wheel. Not surprisingly, in systems with many degrees of freedom, the accompanying phase space is large, and we can expect chance to play an important role in determining how the system evolves. Even in systems with a few degrees of freedom, such as those associated with games of chance governed by deterministic equations of motion, the outcomes can be random due to the amplification of small variations in the initial conditions. Poincar\'e was the first to think physically about probability in his classic paper on the roulette wheel.\cite{Poincare} Later Hopf showed that the underlying physical processes are responsible for the regularity property of probability;\cite{Hopf} that is, the observed frequencies are almost constant. This advance was particularly important because it provided a physical context for the method of arbitrary functions introduced by Poinca\'re { in determining probability distributions.} Hopf was among the first to show the role of the laws of physics in determining the flow of initial distributions to the final states in a system that can be described in terms of probability distributions, particularly when considerations of symmetry or invariance alone do not suffice.\cite{Engels, von Plato}

In the context of the coin toss, the purportedly random outcome can be understood in different ways. Statistically, the equal likelihood of heads and tails is suggested from an analysis of a large sequence of experiments that sample the space of outcomes. An interpretation based on symmetry suggests that because there are only two possibilities (for a coin of zero thickness), both faces should have equal probabilities. This conclusion assumes that the coin can actually explore both configurations (heads and tails ) with equal likelihood, which is not always true in a real coin toss. { For example, a coin that does not flip, but precesses as it spins can end up the same way as it started. }To really understand the randomness in the outcome of a coin toss, we must introduce probability into a mathematical and physical description of the process. { A distribution of initial conditions evolves dynamically leading to outcomes that have effectively ``forgotten" the initial conditions either because the system parameters and/or initial conditions take on particular ranges of values and/or the system has extreme sensitivity to initial conditions (such as in chaotic systems with one or more positive Lyapunov exponents). }

In this paper we focus on the simplest situation corresponding to the case for which the system has dynamical equations that are integrable and allow us to explicitly understand how a distribution of the initial conditions leads to effectively random outcomes via a dynamical flow. A first step in incorporating probability into the physics of the coin toss was done by Keller,\cite{Keller} who considered the simple but illuminating case of a coin of zero thickness spinning about a horizontal axis passing through a diameter, which eventually lands without bouncing. He showed that such a coin toss becomes fair, that is, $P(\mbox{heads}) = P(\mbox{tails}) = 1/2$, in the asymptotic limit of infinite angular velocity {$\omega$} and vertical velocity $u$, when the phase space of any probability distribution about some nominally deterministic initial conditions $(\omega, u)$ is homogeneously and equally divided between the possible outcomes, that is, heads and tails. He showed by explicit calculation how the flow of the dynamical system with a distribution of initial conditions to the final outcome determines the probability of the outcomes. A later report included the dynamics of bouncing in the plane into this minimal model and showed how any initial probability distribution is whittled away exponentially fast.\cite{Keller1}

Adding a third dimension involves a number of new effects -- the coin has two more rotational degrees of freedom in addition to one translational degree of freedom (which is irrelevant); the complex dynamics of bouncing because the coin can land on its edge, side, or face and thus end up neither with heads or tails; and the finite thickness of the coin, which is more cylinder-like. Diaconis, Holmes, and Montgomery analyzed the three-dimensional dynamics of the toss of a coin of zero thickness,\cite{Diaconis} and emphasized the role of the bias induced by the initial conditions. Others have studied the effects of bouncing on a substrate to understand how collisions can also lead to randomness.\cite{Prange, Bondi, Nagler, Nagler1} A recent book elaborates on { the nature of randomness in mechanical games of chance, including the coin toss, by including} the effects of air resistance and bouncing.\cite{book} These more complex models and experiments serve to confirm that the randomness in a coin toss stems primarily from the dynamical flow that acts on the uncertainty in the initial conditions.

Accounting for the finite thickness of a coin leads to a new possibility and increases the phase space of outcomes to include that of landing on an edge -- an event that has a small but nonzero probability.\cite{Murray} For a cylindrical coin of thickness $h$ and diameter $D=2a$, which is tossed and lands without bouncing, the probability of landing on a side is a function of its aspect ratio $\xi = h/D$. The coin will almost surely land on a face when $\xi \rightarrow 0$, and will almost surely land on its side when $\xi \rightarrow \infty$. Continuity suggests that as $\xi \in [0, \infty)$ is varied, so will the probability of landing on either heads/tails or sides. This variation leads naturally to two related questions. What is the aspect ratio of a fair ``3-sided" coin? A fair 3-sided coin is one that starts with a vigorous initial spin and large upward velocity, and lands on heads, sides, and tails with equal probability. How can we build coins with a prescribed probability for landing on their side or face?

Mosteller\cite{Mosteller} described an anecdote about how John von Neumann solved the problem of a fair 3-sided coin almost as it was posed, announcing the answer to an astonished audience, ``$\xi = 1/2 \sqrt 2 \approx 0.357$!" von Neumann must have solved this problem using considerations of symmetry and the geometrical notion of fairness; that is, assuming all possible orientations of the coin are likely, what proportions should the disk-like coin have so that the areal projection of its faces and sides on a circumscribing sphere are identical to each other? Although this assumption is plausible for a rapidly spinning coin, it neglects that the spinning coin must satisfy Newton's equations of motion (actually, Euler's equation for rigid body dynamics) and this equation enforces some conservation laws (angular momentum in particular). This system also highlights a classical conundrum in probability known as ``Bertrand's paradox." That is, the probabilities of an event are ill defined unless the mechanism that produces the random variable is clearly prescribed.\cite{Bertrand} A way around Bertrand's paradox is to use the principle of ``maximum ignorance,'' as given by Jaynes,\cite{Jaynes} and then von Neumann's result is correct. Given the knowledge of a physical law, we must account for it, and we cannot ignore the conservation of angular momentum.

In this paper we use the geometry and dynamics of rigid body motion to derive simple analytical expressions for the probability of landing on heads, sides, or tails for a coin that is tossed vigorously, spins in the air, and lands without bouncing on an inelastic substrate, such as the palm of one's hand or a pile of sand. These expressions generalize the earlier results for coins of {zero thickness, \cite{Diaconis}} and allow us to see how probability depends on the geometry of the coin via its aspect ratio $\xi$ and the dynamical angle $\psi$ which characterizes the precession of the coin, as determined by its initial angular momentum. We find that a notion of fairness based on rigid body dynamics yields a fundamentally different probability distribution for the outcomes compared to the result based on the purely symmetry-based notion of fairness. In particular, the new criterion yields an aspect ratio of $\xi = 1/\sqrt 3$ for an equal probability of heads, sides, and tails when {the coins are spun rapidly.} \cite{Maha11}

Simple experiments qualitatively confirm our theory and allow us to prescribe criteria for designing coins with a prescribed probability distribution of landing on heads, tails, or sides. Our approach also allow us to illustrate the role of skill as exemplified by the ability to bias the outcome of the coin toss using the law of conditional probabilities. 

\section{Dynamics of spin}

\subsection{Mathematical formulation}

\begin{figure}[h!]
\centering
\includegraphics[width=6in]{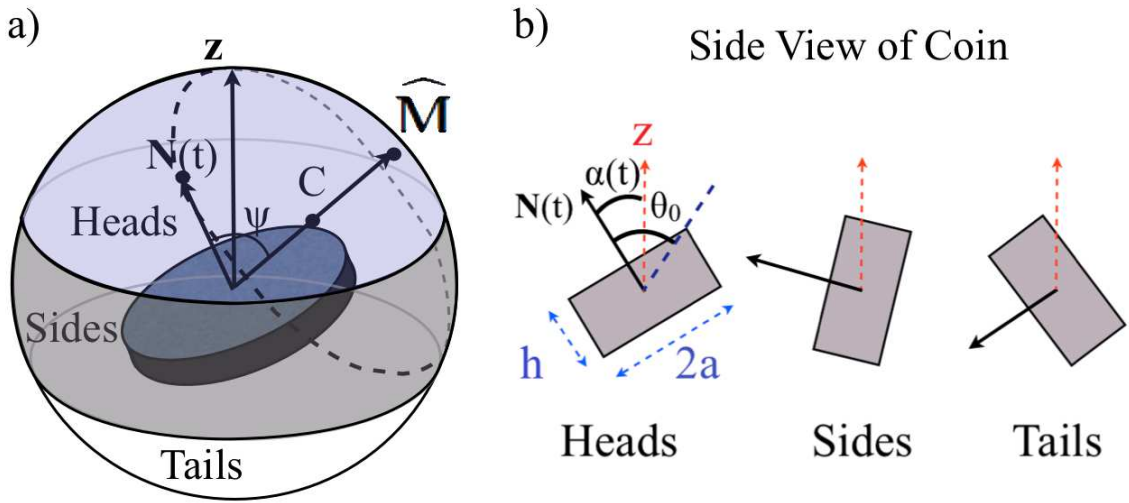}
\caption{ (a) For a spinning, precessing coin with its normal pointing vertically upward at time $t=0$, that is, $\mathbf{N}(0)=\mathbf{z}$, conservation of angular momentum dictates that $\mathbf{N}(t)$ sweep out a (dotted) circle on the unit sphere. (b) If the coin lands without bouncing, the side which faces up is determined by the difference between the dynamical angle $\alpha(t_f)$ and the static angle $\theta_0= \cos^{-1} (\xi/\sqrt{1+\xi^2})$, where $\xi = h/2a$ is the aspect ratio of the coin. Thus, we obtain heads if $ 0\le \alpha_f \le \theta_0$, sides if $ \theta_0 < \alpha_f \le \pi - \theta_0$ and tails if $\pi - \theta_0 < \alpha_f \le \pi$.}
\label{sphere}
\end{figure}

We assume that a coin is made of a homogeneous material and is axisymmetric, with its initial orientation such that the normal vector $\mathbf{N}(t)$ outward from the head points upward, that is, $\mathbf{N}(0) = \mathbf{z}$, and its initial angular velocity is $\mathbf{\Omega}$. Therefore its angular momentum is $\mathbf{M}= \mathbf{I}\mathbf{\Omega}$, where $\mathbf{I}$ is the moment of inertia tensor, with principal moments of inertia $I_1 = I_2 = \frac{1}{4} ( ma^2 + \frac{1}{3} m h^2)$ and $I_3 = \frac{1}{2}ma^2$, so that $\mathbf{M} = I_1 \mathbf{\Omega} + (I_3 - I_1) \omega_3 \mathbf{N}$. We write the angular momentum relative to a lab-fixed frame $\mathbf{X}(t)$ that is connected to the body-fixed frame $\bf{x}$ by the relation ${\bf X}(t) = {\bf Q}(t) {\bf x}$, with the rotation matrix ${\bf Q}(t) \in \mbox{SO}(3)$, so that the body-fixed angular velocity $\boldsymbol{\omega} = {\bf Q}^T(t) {\mathbf \Omega}$, and the body-fixed angular momentum ${\bf m} = {\bf Q}^T(t) {\bf M}$. The evolution of the unit normal to the coin is given by\cite{Landau}
\begin{equation}
\frac{d \mathbf{N}}{dt} = \boldsymbol{\Omega} \times \mathbf{N}.
\label{mech1}
\end{equation}

If $\psi$ is the angle between the angular momentum $\mathbf M$ and $\mathbf{N}(t)$ at time $t = 0$ given by $\cos(\psi) = \mathbf{N}(0)\cdot \widehat{\mathbf{M}}$ where $\widehat{\mathbf{M}} = \mathbf{M}/ M$, $M=||\mathbf{M}||$, and $\omega_N = M/I_1$, we find that
\begin{equation}
\frac{d \mathbf{N}}{dt} = \omega_N \widehat{\mathbf{M}} \times \mathbf{N}.
\label{mech2}
\end{equation}
That is, the normal to the coin sweeps out a cone as it precesses about the axis $\widehat{\mathbf{M}}$ with the frequency $\omega_N$, keeping the angle between the angular momentum vector and the normal $\psi$ constant for all time. On the unit sphere, $\mathbf{N}(t)$ traces a circle which contains the ``north pole'' ($\mathbf{z}$) as shown in Fig.~\ref{sphere}(a). The projection of the normal in the up direction $\mathbf{z}$ is\cite{Diaconis}
\begin{equation}
f(t) = \mathbf{N}(t)\cdot\mathbf{z} = \cos \alpha(t) = A + B \cos \theta(t),
\label{angle}
\end{equation}
where $A=\cos^2 \psi, B = \sin^2 \psi$ and $\theta(t)=\omega_N t$.

\subsection{Heads, sides or tails?}

\begin{figure}[h!]
\centering
\includegraphics[width=6in]{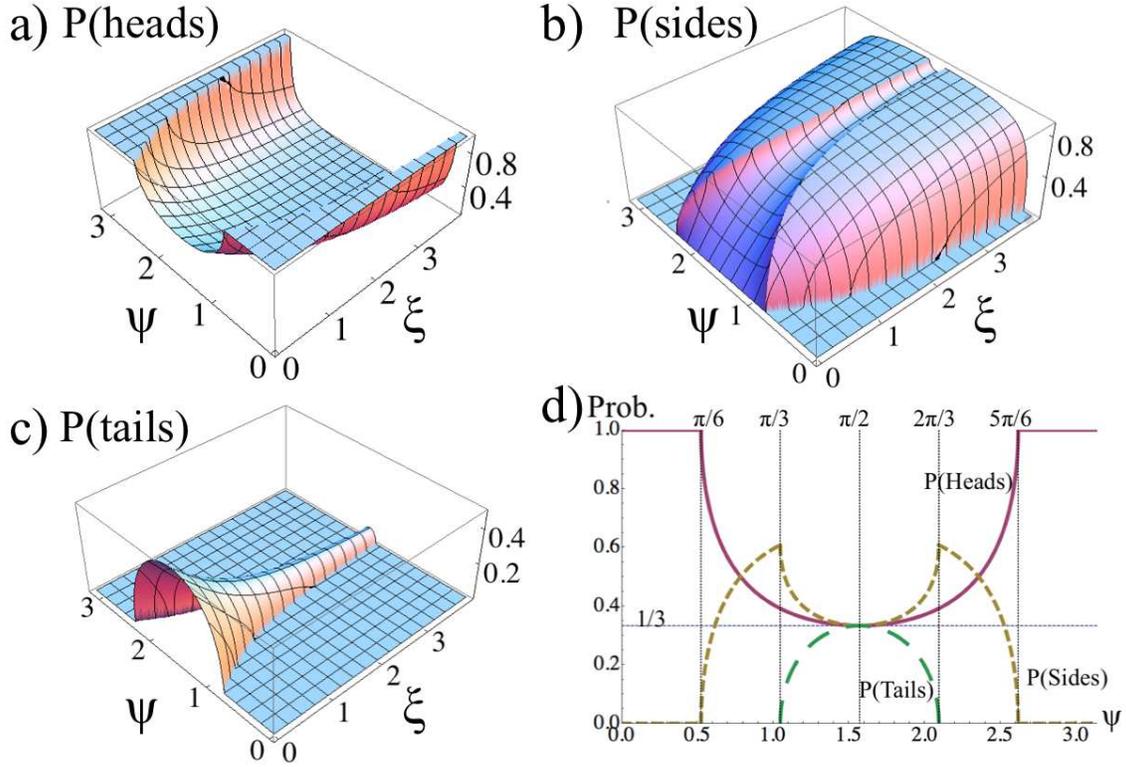}
\caption{Probability distribution of landing on heads, sides, and tails as a function of the angle $\psi$ between the the angular momentum vector $\mathbf M$ and the normal to the coin $\mathbf N$, defined by $\cos \psi = \mathbf{N}(0)\cdot\widehat{\mathbf{M}}$ and the aspect ratio of the coin $\xi= h/D$. (a) $P$(heads) as a function of $\xi$ and $\psi$, (b) $P$(sides), and (c) $P$(tails). (d) A section through the figures for $\xi = 1/\sqrt{3}$ shows the probability distribution of landing on heads (solid curve), sides (small dashed) and tails (long dashed) as a function of $\psi$. Only for $\psi=\pi/2$ is the coin dynamically fair so that it is equally likely to land with heads, tails, or sides up when tossed vigorously with $\omega_N t_f \gg 1$.}
\label{result}
\end{figure}

When such a coin falls onto a substrate without bouncing, its normal vector $ \mathbf{N}(t)$ at that instant determines whether the coin lands on its heads, sides, or tails, depending on the difference between the dynamical angle $\alpha_f = \alpha(t_f)= \cos^{-1}( \mathbf{N}(t_f) \cdot \mathbf{z})$ given by Eq.~(3) and the angle $\theta_0 = \cos^{-1} (\xi/\sqrt{1+\xi^2})$ which the diagonal to the coin makes with the normal [see Fig.~\ref{sphere}(b)]. The time of flight, $t_f$, can be found by solving Newton's equations for the center of mass of the coin:
\begin{equation}
\frac{d^2 z(t)}{d t^2} = -g, \quad z(0) = \frac{\sqrt{3}}{2}a, \quad \frac{dz(0)}{dt} = u.
\end{equation}
{where the particular choice of $z(0)$ simplifies some of the subsequent calculations.} If the coin is caught at height $z = 0$, then $t_f$ is the smallest positive root of the equation [see Fig.~\ref{sphere}(b)]
\begin{equation}
z(t_f) - a \sin \alpha (t_f) = 0,
\end{equation}
and the criteria for landing on heads, sides, and tails are respectively given by
\begin{subequations}
\label{cutoff}
\begin{align}
& 0 \le \alpha_f \le \theta_0, && \text{heads} \\
& \theta_0 < \alpha_f \le \pi - \theta_0, && \text{sides} \\
& \pi - \theta_0 < \alpha_f \le \pi, && \text{tails},
\end{align}
\end{subequations}
which divides the surface of the unit sphere into three zones: a polar spherical cap for heads, a middle equatorial zone for sides, and another polar spherical cap for tails as shown in Fig.~\ref{sphere}(a).

\textit{Problem 1}. Instead of a coin, suppose we toss a book into the air. In this case the principal moments of the book, a rectangular prism, are all different ($I_3 > I_2 > I_1$, non axi-symmetric), and hence Eq.~(\ref{mech2}) does not apply, and we have to resort to Euler's equations. At time $t=0$ the book is flipped with angular velocity $\boldsymbol{\omega}(0) = (0, 0, \omega_0)$, where $\omega_0 \gg 1$. Show that $\omega_3(t)$ is approximately constant throughout the motion, and that in this case $\omega_1(t)$ and $\omega_2(t)$ are bounded and oscillate with frequency $\Gamma$ given by
\begin{equation}
\Gamma^2 = \frac{(I_3-I_1)(I_3 - I_2)}{I_1 I_2} \omega_0^2.
\end{equation}
If we start off with $\boldsymbol{\omega}(0) = (0, \omega_0, 0)$, where $ \omega_0 \gg 1$, then $\omega_1(t)$ and $\omega_3(t)$ do not remain small. Analyze this case. {The axisymmetry of the coin provides a great deal of simplification. For a polyhedral dice toss, we have to track the vertical velocity, the angular velocity vector $\boldsymbol{\Omega}$, and the evolution of the body-fixed frame $\bf{x}$, which makes the problem more complicated, but worthy of study.}

\section{Dynamics, probability and geometry}

\subsection{The general case}

To link the physics of spin and precession to probability, we consider the phase space of initial conditions. Because coins are usually flipped vigorously, we might imagine that the angle associated with the spin is uniformly distributed. This assumption does hold as shown by Kemperman and Engel,\cite{Engels} who proved that for vigorously flipped coins, i.e. $\omega_N u/g \gg 1$, { where $\omega_N$ is the precessional frequency as defined earlier and $u$ is the magnitude of the upward velocity}, the quantity $\theta_f = \omega_N t_f$, modulo 2$\pi$, approaches a uniform distribution on the interval $[0, 2\pi)$. Because the function $f(\theta)$ in Eq.~(\ref{angle}) is symmetric about $\theta = \pi$ and monotonically decreasing on $(0, \pi)$, it follows that there is a unique value of $\theta_1$ in $(0, \pi)$ that defines the landing condition $f(\theta_1) = \cos \alpha_f= A + B \cos \theta_1 = \cos \theta_0$, where $A = \cos^2 \psi$ and $B = \sin^2 \psi$. Thus, the probability of heads, $P(\text{heads})$, given by the uniform measure of the set $\{ \mbox{$\theta$:}\, f(\theta) > \cos \theta_0 \}$, is
\begin{equation}
P(\text{heads}) = \theta_1/\pi.
\end{equation}

We can now calculate the full probability distribution for a coin with arbitrary aspect ratio $\xi = h/D$, that is, $\xi \in [0, \infty)$, and arbitrary angular momentum vector $\mathbf{M}$, that is, $\psi \in [0, \pi]$. Because $A - B = \cos^2 \psi - \sin^2 \psi = \cos(2\psi)$, P(heads) $=1$ when $1 \ge A - B > \cos \theta_0$ so that $\psi \in [0, \theta_0/2) \cup (\pi - \theta_0/2, \pi]$, and the normal to the coin precesses about the angular momentum vector making an angle in the range $(0, \psi)$ relative to the vertical axis. Similarly if $\cos \theta_0 \ge A - B \ge -\cos\theta_0$, that is, $\psi \in [\theta_0/2, \pi/2 -\theta_0/2) \cup (\pi/2 + \theta_0/2, \pi - \theta_0/2]$, the coin will only land on heads or sides, and if $-\cos \theta_0 > A - B \ge -1$, that is, $\psi \in [\pi/2 - \theta_0/2, \pi/2 + \theta_0/2]$, the coin can land on heads, sides, or tails. A geometrical way of understanding this result follows by tracking the trajectory of the tip of the unit normal vector $\mathbf{N}(t)$, which traces three possible distinct classes of circles: a circle that lies entirely in the polar heads zone, a circle that lies in the polar-equatorial heads and sides zone, and a circle that lies in all three zones. We define
\begin{align}
\theta_1 &= \cos^{-1}\left( \frac{\cos\theta_0 - \cos^2 \psi}{\sin^2 \psi} \right)\\
\noalign{\noindent and}
\theta_2 &= \cos^{-1}\left( \frac{-\cos\theta_0 - \cos^2 \psi}{\sin^2 \psi} \right),
\label{angles}
\end{align}
and obtain the three types of solutions as shown in Table~I.

In Fig.~\ref{result} we plot the probability distribution for landing on heads, sides, and tails as a function of $\xi$ and $\psi$ in the limit of high spin. These results complement the earlier results for the planar flip of a coin of zero thickness,of Keller\cite{Keller} $\xi = 0$, $\psi = \pi/2$ and the three-dimensional dynamics of a coin of zero thickness: the line $\xi = 0,\psi \in [0, \pi]$.\cite{Diaconis} As expected, we see that vigorously tossed thick coins that start heads-up are biased to come heads-up because there is a large range for the initial angle $\psi$ that favors this outcome.

\subsection{The dynamically fair coin}

A dynamically fair coin is one where $P(\text{heads})=P(\text{sides})=P(\text{tails})=1/3$ (see Table~I), so that $\theta_2=2\pi/3 = 2\theta_1$ and $\psi = \pi/2$, $\cos\theta_0 =1/2$, and the aspect ratio of the coin $\xi =1/\sqrt{3}$, in contrast with the condition for a geometrically fair coin, where $\xi = 1/2 \sqrt{2}$.\cite{Mosteller} Only for this unique combination of coin geometry and orientation of the angular momentum vector $\{\xi, \psi\}$, does the trajectory of the unit normal vector $\mathbf{N}(t)$ transverse a great circle containing the meridian (line of longitude) on the unit sphere with equal length of the trajectory in the heads, sides, and tails regions. This point is the only one in the phase space of $\{\xi, \psi\}$ for which there are equal probabilities for heads, sides, and tails in the dynamical sense. Thus, we can only obtain a fair result when tossing a thick coin under the ``Keller flip'' condition: {coin starts heads up with $\psi = \pi/2$.}

\begin{figure}[h!]
\centering
\includegraphics[width=6in]{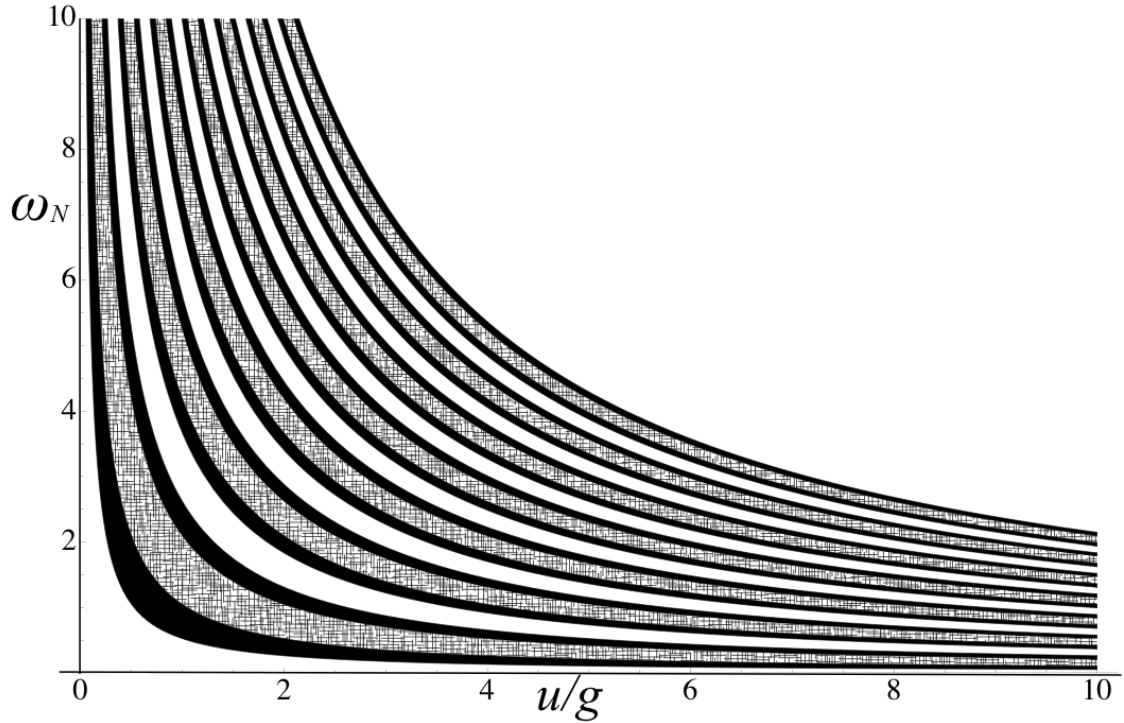}
\caption{Phase space of possibilities for a thick coin. For different aspect ratios $\xi$, hyperbolae separate the phase space into regions of heads (white regions), sides (black regions) and tails (stripes regions) as defined in Eq.~(\ref{omega}).  The case shown corresponds to the fair coin when $\xi = 1/\sqrt{3}$ and $\psi = \pi/2$, and shows that sides appear twice as often, but with half the area associated with heads and tails. Far from the origin, corresponding to arbitrarily large values of $u$ and $\omega$ and a vigorously spun coin, any disk of arbitrarily small area will contain equal proportions of heads, sides, and tails regions, because the hyperbolae become more closely spaced, and approaches the limiting case of a fair 3-sided coin. }
\label{phase}
\end{figure}

For $\xi = 1/\sqrt{3}$ we find that the coin always lands heads up when $1 \ge A - B > 1/2$; that is, $\psi \in [0, \pi/6) \cup (5\pi/6, \pi]$. The coin only land son either heads or sides when $1/2 \ge A - B \ge -1/2$; that is, $\psi \in [\pi/6, \pi/3) \cup (2\pi/3, 5\pi/6]$. The coin can land on either heads, sides, or tails when $-1/2 > A - B \ge -1$; that is, $\psi \in [\pi/3, 2\pi/3]$. Thus, we have the following three cases:
\begin{subequations}
\begin{align}
P(\text{heads}) &= 1, \,P(\text{sides}) = P(\text{tails}) = 0, && \psi \in [0, \pi/6) \cup (5\pi/6, \pi] \\ 
P(\text{heads}) &= \dfrac{\theta_1}{\pi}, \,P(\text{sides}) = \dfrac{\pi - \theta_1}{\pi},\; P(\text{tails}) = 0, && \psi \in [\pi/6, \pi/3) \cup (2\pi/3, 5\pi/6]  \\ 
P(\text{heads}) &= \dfrac{\theta_1}{\pi}, \,P(\text{sides}) = \dfrac{\theta_2 - \theta_1}{\pi}, \;P(\text{tails}) = \dfrac{\pi-\theta_2}{\pi}, && \psi \in [\pi/3, 2\pi/3].
\end{align}
\label{main}
\end{subequations}
The probability outcomes $P(\text{heads})$, $P(\text{sides})$, and $P(\text{tails})$ as a function of $\psi$, the angle between the normal of the coin to the angular momentum vector are plotted in Fig.~\ref{result}.

\subsection{Phase space of pre-images of a thick tossed coin}

To understand how the probability distribution of initial conditions evolves through the flow and leads to random outcomes, we consider how the phase space of possibilities, that is, heads, sides, or tails, is mapped onto the initial conditions, that is, the pre-images, which lead to these different outcomes. In Fig.~\ref{phase} we show the pre-images of heads, sides, and tails of a dynamically fair coin, that is, one tossed upward with $\psi = \pi/2$ and $h/D = 1/\sqrt{3}$. After the coin has landed (without bouncing), it has rotated $\omega_N t_f$ times. Depending on the number of revolutions $n$, where $n$ is an integer, the coin lands on its head if $2 n \pi \pm \theta_0 = 2 n \pi \pm \pi/3 = \omega_N t_f$ and lands on its tail if $2 (n+1) \pi \pm \pi/3 = \omega_N t_f$; otherwise the coin lands on its sides. Thus the phase space $(\omega_N, t_f)$ may be decomposed into the regions with boundaries of the regions given by the hyperbolae
\begin{subequations}
\begin{align}
\omega_N &= \left( 2n \pm \frac{1}{3} \right) \frac{\pi g}{2u} && n=0,1,2,\ldots \; \text{(heads)} \\
\omega_N &= \left( (2n+1) \pm \frac{1} {3} \right) \frac{\pi g}{2u}, && n=0,1,2,\ldots \; \text{(tails)}.
\end{align}
\label{omega}
\end{subequations}
On the axis $\omega_N = 0$, the coin remains heads up throughout the toss, and therefore this axis and the adjacent strip lie in H, the pre-image of heads {as shown in Fig.~\ref{phase}.} The next strip lies in S, the pre-images of sides; the next strip lies in T, the pre-image of tails; the next strip lies in S, and the sequence H, S, T, S repeats itself. We see that the hyperbolae striate phase space ever more finely as the spin $\omega_N$ and the scaled velocity $u/g$ increase. Each region of H and T have equal area while S is half as large but occurs twice as often. As we shift a finite area disk in this phase space to infinity, we find that H, S, and T occupy fixed and equal areas of the disk, so that the coin toss becomes dynamically fair only asymptotically.

\begin{figure}[h!]
\centering
\includegraphics[width=6in]{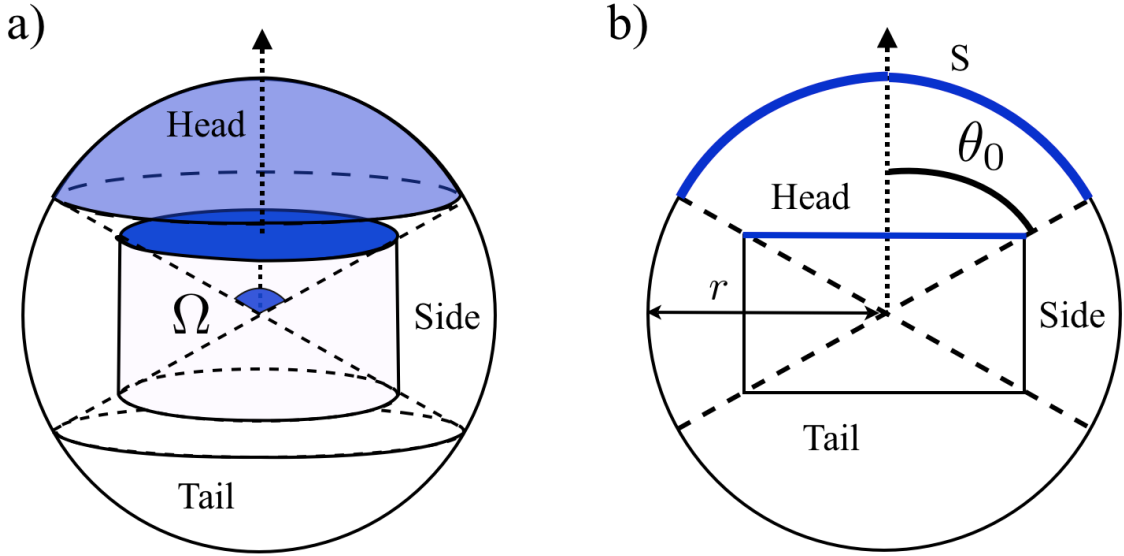}
\caption{(a) The purely symmetry-based argument of von Neumann, \cite{Mosteller} connects probability to the (assumed) uniform distribution of orientations in space, so that the probability of heads is the ratio of the solid angle $\Omega_s$ subtended by the head of the coin to the total solid angle of a unit sphere, that is, $\Omega_s/4\pi$. For a fair coin with an equal probability of landing on its head, tail, or side, $\Omega_s=4\pi/3$ and $\xi = 1/2\sqrt{2}$. (b) For the dynamical argument associated with the Keller flip (the only case where for a vigorous flip, it is possible to eliminate the bias based on initial conditions), the probability of heads is the ratio of the arc length $s$ subtended by heads and the circumference of the circle, that is, $s/(2\pi r) = \theta_0/\pi$, so that for a fair 3-sided coin, $\theta_0 = \pi/3$ and $\xi=1/\sqrt{3}$.}
\label{geom}
\end{figure}

\subsection{A geometrical view}

As we have seen, the basic difference between the von Neumann flip and the Keller flip can be characterized in terms of the geometry of allowable orientations. A minimal view of von-Neumann's argument is shown in Fig.~\ref{geom}(a). Given the assumption of a uniform distribution of all possible orientations, the probability of landing on heads and tails is given by the ratio of the solid angle subtended by heads (or tails) $\Omega_s$ to the total solid angle of a unit sphere, that is, $\Omega_s/4\pi$, and that of landing on a side is $1-2\,\Omega_s/4\pi$. Therefore, a fair 3-sided coin must be such that $\Omega_s/4 \pi = 1/3$. If we use the usual spherical coordinate system, we can calculate $\Omega_s = 2\pi(1-\cos \theta)$, where $\theta$ is the half-angle subtended by the face. If we equate the two relations, we find that $\cos \theta = 1/3$. By definition, $\cos \theta = \xi/\sqrt{1+\xi^2}$, so that the aspect ratio of the coin is $\xi = 1/2\sqrt{2}$.

In contrast, the constraint of constant angular momentum leads us to consider the { Keller flip}, the only truly unbiased flip. In this case we consider a projection of a cross-section of the coin onto a circumscribed circle, as shown in Fig.~\ref{geom}. The probability of landing on a particular face (or side) is now the ratio of the arc subtended by the face (or side) divided by the entire circle ($2\pi$). Thus, $P(\mbox{heads}) = \theta_0/\pi$, $P(\mbox{tails}) = \theta_0/\pi$, and $P(\mbox{sides}) = 1 - P(\mbox{heads}) - P(\mbox{tails}) = 1 - 2\theta_0/\pi$, so that for a fair 3-sided coin, $\theta_0 = \pi/3$, and thus the aspect ratio of the coin $\xi= 1/\sqrt{3}$.

Thus, we see another example of how Bertrand's paradox arises naturally. Depending on the assumptions of the mechanism (or equivalently, the implied symmetry and invariance) which produces the random variable, the probabilities are ill-defined and thus lead to different answers for the aspect ratio of a fair coin.

\section{Experiments}

\begin{figure}[h!]
\centering
\includegraphics[width=6in]{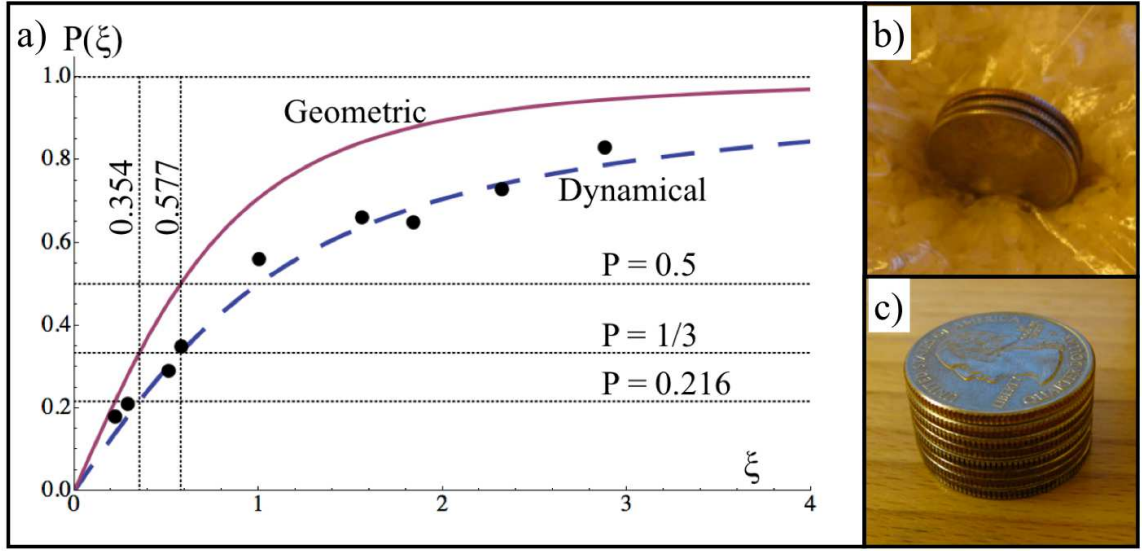}
\caption{(a) Probability of sides for a vigorously spun coin ($u \omega/g \ge 20$)   as a function of the aspect ratio $\xi$. The dots correspond to our experiments and denote the frequency of sides for 100 flips. The solid and dashed lines corresponds to the geometrical and dynamical definitions in the text. (b) The probability of a 3-coin landing on sides on an highly inelastic surface made of rice grains. (c) A dynamically fair 3-sided coin composed of stacking 8 U.S.\ quarters, with an aspect ratio $\xi \approx 0.58$.}
\label{scale}
\end{figure}

To test our theoretical results, we conducted a series of simple tabletop experiments. To make thick coins, we glued U.S.\ quarters, of diameter 24\,mm and thickness 1.75\,mm, together to form an $N$-coin of different aspect ratios, for example, a 3-coin is formed by gluing three U.S.\ quarters together, and tossed them by hand with $\psi \sim \pi/2$ and starting heads up, onto a highly inelastic surface, such as a pan of rice covered by a thin film of plastic. For thicker coins, we cut cylindrical pieces of an aluminum rod of diameter 25\,mm. Each coin was vigorously tossed ($u \omega/g \ge 20$) 100 times starting with heads up, and the experimentally determined frequency of sides is plotted (as dots) in Fig.~\ref{scale}. The sum of squared errors for the geometrical case ($0.20$) is significantly larger than for the dynamical case ($0.01$). Our experimental results are in good agreement with the predictions of the dynamical theory, and suggest a simple criterion for the aspect ratio of designer coins with a given bias to land on a side or a face.

\begin{table}[h!]
\centering
\begin{tabular}{|c|c|c|c|}
\hline
Range of $\psi$ & P(heads) & P(sides) & P(tails) \\
\hline
$[0, \theta_0/2) \cup (\pi - \theta_0/2, \pi]$ & 1 & 0 & 0\\
\hline
$[\theta_0/2, \pi/2 -\theta_0/2) \cup (\pi/2 + \theta_0/2, \pi - \theta_0/2]$ & $\theta_1/\pi$ & $1 - \theta_1/\pi$ & 0 \\
\hline
$[\pi/2 - \theta_0/2, \pi/2 + \theta_0/2]$ & $\theta_1/\pi$ & $(\theta_2 - \theta_1)/\pi$ & $1 - \theta_2/\pi$ \\
\hline
\end{tabular}
\caption{Probabilities of heads, sides, and tails  for a coin of arbitrary aspect ratio $\xi$ and different values of $\psi$.} 
\label{solution-table}
\end{table}

It is useful to compare these experimental results with both geometrically and dynamically fair coins in the limit of thin and thick coins. For arbitrary aspect ratio $\xi$, the symmetry-based geometrical view implies that the probability of landing on sides is given by
\begin{equation}
P_G(\text{sides}) = 1 - 2P(\text{heads}) = 1 - \frac{1}{2}\!\int_0^{\theta_0} \sin \theta \,d \theta = \cos \theta_0 = \frac{\xi}{\sqrt{1+\xi^2}}.
\label{geometric}
\end{equation}
A geometrical fair coin has $\xi = 1/2\sqrt{2}$, which implies that $P_G(\text{sides}) = 1/3$, and a dynamical fair coin ($\psi = \pi/2$) has $\xi = 1/\sqrt{3}$, which under the geometrical prediction, gives a probability of $P_G(\text{sides}) = 1/2$.

In contrast, using the dynamical view that respects conservation of angular momentum, the probability of landing on a side is given by
\begin{equation}
P_D(\text{sides}) = \frac{\pi -2\theta_0}{\pi} = 1 - \frac{2}{\pi} \cos^{-1}\left( \frac{\xi}{\sqrt{1+\xi^2}} \right).
\label{keller}
\end{equation}
Therefore $\xi = 1/\sqrt{3}$, $P_D(\text{sides}) = 1/3$, and for $\xi = 1/2\sqrt{2}$, $P_D(\text{sides}) = 0.216$. In the small $\xi$ limit we find
\begin{subequations}
\begin{align}
P_G(\text{sides}) &= \xi - \frac{\xi^3}{2} + O(\xi^4)\\
\noalign{\noindent and}
P_D(\text{sides}) & = \frac{2}{\pi} \xi - \frac{2}{3\pi}\xi^3 + O(\xi^5).
\end{align}
\end{subequations}
In contrast, in the large $\xi$ limit, we find
\begin{subequations}
\begin{align}
P_G(\text{sides}) & = 1 - \frac{1}{2\xi^2} + O\left(\frac{1}{\xi^4}\right) \\
\noalign{\noindent and}
P_D(\text{sides}) & = 1 - \frac{2}{\pi \xi} + O\left(\frac{1}{\xi^2}\right).
\end{align}
\end{subequations}
Although a coin with vanishing thickness has vanishing probability of landing on its side and an infinitely long coin will always land on its sides, we find that $P_G(\text{sides})$ approaches the asymptotes at a much faster rate than $P_D(\text{sides})$ as shown in Fig.~\ref{scale}(a).

\textit{Problem 2}. Use Eq.~(\ref{keller}) and write a Monte Carlo program that can find the dynamical probability of landing on the sides of a thick coin for a given $\xi$. Generalize your program to consider a coin with arbitrary angular momentum vector. For the case for which $\psi$ has a normal distribution with mean $\pi/2$ and variance $0.1$, show that this distribution results in a curve that is slightly displaced above the (dashed) dynamical curve in Fig.~\ref{scale}(a).

\section{Discussion}

By adding the thickness dimension of a coin, we have expanded the phase space of possibilities of a coin toss landing on an inelastic substrate and derived simple expressions for the probability of landing on a side as a function of the aspect ratio of the coin and its initial orientation relative to its angular momentum vector. { Our simple model} allowed us to derive the conditions for a dynamically fair 3-sided coin: we must toss a coin of aspect ratio $h/D=1/\sqrt{3}$ with its angular momentum lying in its plane. that is, $\psi = \pi/2$, just as for a coin of zero thickness.\cite{Diaconis} We also saw how the coin toss is a natural example of Bertrand's paradox and its resolution using physical principles (embodied in terms of symmetry and invariance) which have a direct geometrical interpretation.

\begin{table}[h!]
\centering
\begin{tabular}{|c|c|c|c|}
\hline
& Uniform & Cosine & Normal \\
\hline
P(heads) & 0.630 & 0.439 & 0.478 \\
\hline
P(sides) & 0.281 & 0.396 & 0.371 \\
\hline
P(tails) & 0.088 & 0.165 & 0.150 \\
\hline
\end{tabular}
\caption{Probabilities of heads, sides, and tails for three distributions for $P(\psi)$ when tossing a coin with heads up. The distributions are $P(\psi)=1/\pi$ (uniform), $P(\psi)=(1 - \cos 2\psi)/\pi$ (cosine), and
$P(\psi)=0.58 \exp^{-(\psi-\pi/2)^2}$ (normal).}
\label{psi}
\end{table}

We conclude with a brief remark on the role of the distribution of $\psi$, the angular variable that describes the relative orientation of the coin normal to the angular momentum vector. As $\psi$ deviates from $\pi/2$, the probability of sides is no longer 1/3. Then $P(i)$, where $i =$ heads, sides, or tails is given by
\begin{equation}
P(i) = \!\int_0^\pi P(i| \psi) P(\psi) d\psi,
\end{equation}
where the conditional probability $P(i|\psi)$ is now given by {Eq.~(\ref{main}).} In Table~II we show the affect of three symmetrical distributions for $P(\psi)$, $\psi \in [0, \pi]$. We consider $P(\psi)=1/\pi$; $P(\psi)=(1 - \cos 2\psi)/\pi$, and $P(\psi)= a \exp^{-(\psi-\pi/2)^2}$ with $a=0.58$. In each case we find that the coin is biased toward heads (the initial condition). This bias suggests learning strategies for novices to become experts and approach the mythical Rosencrantz and the real Diaconis who are able to exploit these deviations to effect long streaks of heads.\cite{Stoppard}

\begin{acknowledgments}
We are grateful to the referees for their detailed comments and especially one for insisting that we emphasize the geometrical underpinnings of the problem throughout and provide Fig.~\ref{geom}(b).
\end{acknowledgments}

\end{document}